\documentstyle[11pt,newpasp,twoside,epsf]{article}
\def\edcomment#1{\iffalse\marginpar{\raggedright\sl#1\/}\else\relax\fi}
\reversemarginpar

\begin{document}
\title{Reversing type II migration: resonance trapping of a lighter giant protoplanet}
 \author{F.S. Masset}
\affil{Astronomy Unit, School of Mathematical Sciences, Queen Mary \& 
Westfield College, 327 Mile End Road, London E1 4NS, UK}
\author{M.D. Snellgrove}
\affil{Astronomy Unit, School of Mathematical Sciences, Queen Mary \& 
Westfield College, 327 Mile End Road, London E1 4NS, UK}

\begin{abstract}

We present new results related to the coupled evolution of a two giant
planet system embedded in a protoplanetary disk, in which a Saturn mass protoplanet is trapped in an
outer mean motion resonance with a Jupiter mass protoplanet. The gaps
opened in the disk by the two planets overlap,
therefore the two planet system exchanges angular momentum with the
disk at the inner planet's Inner Lindblad Resonances (ILRs) and at the
outer planet Outer Lindblad Resonances (OLRs). Since the torques are
proportional
to the square of the planet masses, and since the inner planet is about
three times more massive than the outer one, the ILR torques are
favored by a factor~$\sim 10$ with respect to the one planet case. In
the case presented here, this leads to a positive differential
Lindblad torque and consequently an outwards migration. 
We briefly discuss the long-term behavior of the system,
which could account for the high eccentricities of the extra-solar
planets with semi-major axis $a>0.2$~AU.

\end{abstract}

\section{Introduction}
We consider the coupled evolution of
a system of giant protoplanets embedded in a minimum mass
protoplanetary disk,
consisting of two non-accreting
cores with  masses $1$~$M_J$~and
$0.29$~$M_J$ (where $M_J$ is one Jovian mass), which we call respectively
``Jupiter'' and ``Saturn''. This problem is addressed through numerical
simulations. After a brief description of the codes we have used, we
present
the results of a typical run and discuss several issues.

\section{Numerical codes description and initial setup}
In order to investigate the long-term behavior of the embedded Jupiter and
Saturn system, we used two independent fixed Eulerian grid based
hydrocodes:
NIRVANA (Ziegler \& Yorke, 1997) and
FARGO
(Masset, 2000).
They gave very similar
results. They consist of a
pure N-body kernel embedded in a hydrocode which provides a tidal interaction
with a 2D non self-gravitating gaseous disk. 
In the following our length unit is
$5.2$~AU, the mass unit is one solar mass,
and the time unit is the initial orbital period of Jupiter.
In the run presented here the grid resolution adopted is
$N_r=122$ and $N_\theta=300$ with a geometric spacing of the interzone radii.
The grid outer boundary is at $R_{max}=5$ and its inner boundary is
at $R_{min}=0.4$. 
The protocores
start their
evolution with semi-major axis respectively $a_j=1$ and $a_s=2$.
The disk surface density is 
uniform and corresponds to  two Jupiter masses 
inside Jupiter's orbit. The effective viscosity $\nu$, the nature
of which remains unclear and is usually thought to be due to turbulence generated
by MHD instabilities (Balbus \& Hawley 1991), is assumed to be uniform through
the disk and corresponds to a value of $\alpha \simeq 6\cdot 10^{-3}$ in the vicinity
of Jupiter's orbit. The disk aspect ratio is uniform and constant and
set to $H/r=0.04$.

\section{Run results}
The results are presented in Fig.~1.
The mass of Jupiter is sufficient to open a deep gap
and hence it settles initially in a type II migration (Nelson et
al. 2000) and behaves as if it was the only planet in the disk (see
test run), whereas
Saturn is unable to fully empty its coorbital region, and therefore
starts a much faster migration. 
It reaches the 1:2 resonance with
Jupiter at time 
$t\simeq 110$.
The planets then get higher
eccentricities, and Saturn's migration rate is reduced.
Saturn's eccentricity increases again rapidly as it passes through the 3:5 resonance with
Jupiter at 
$t\simeq 220$,
and eventually 
it
gets trapped into the 2:3 resonance.
Both planets then
steadily migrate {\it outwards}.

\begin{figure}
  \begin{center}
    \plottwo{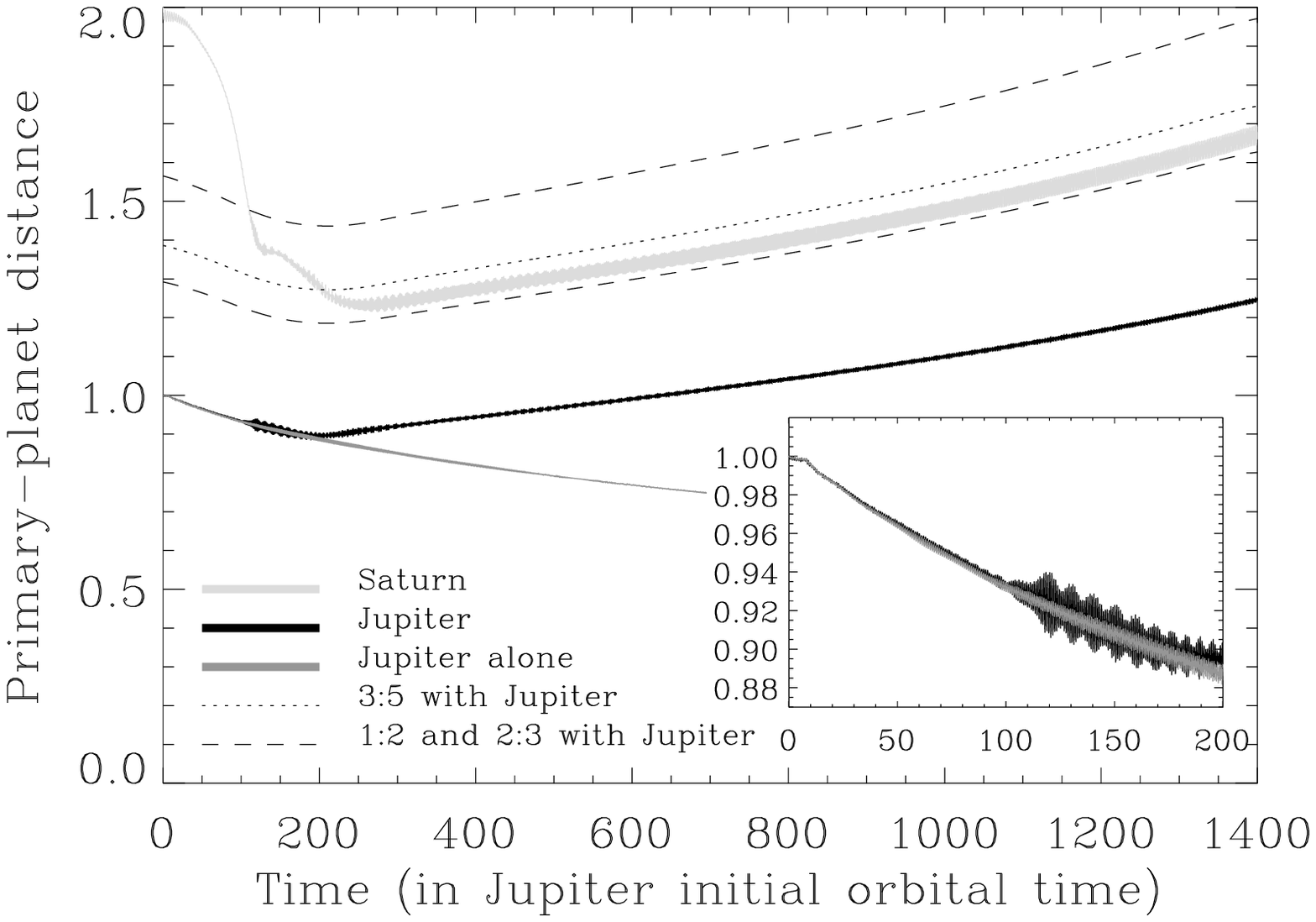}{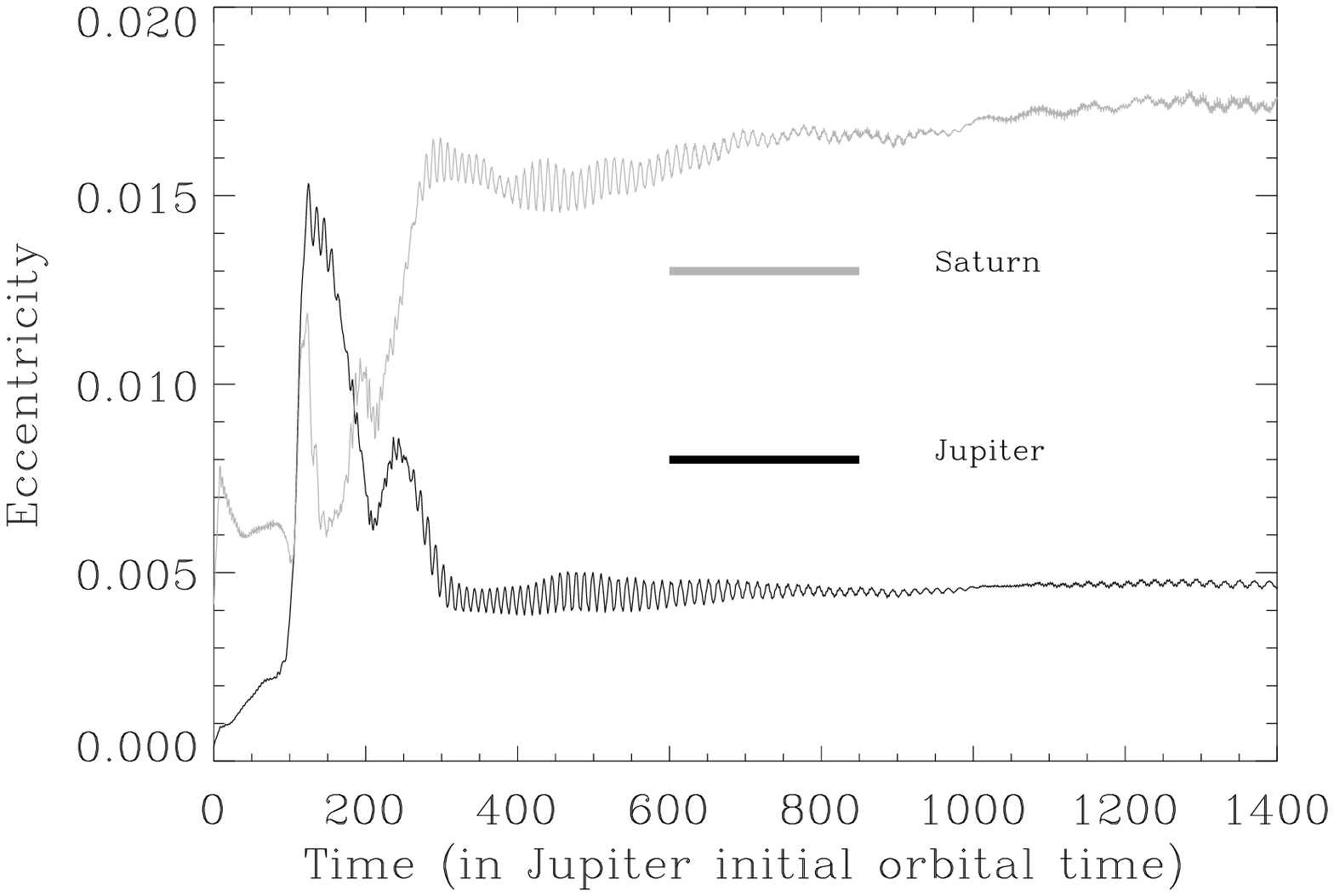}
    \caption{Primary-planet distances  
and planet eccentricities as a function of time. 
On the left plot the medium grey curve shows the results of a 
Jupiter only test run.
}
  \end{center}
\end{figure}

\subsection{Interpretation}
{We define
the system of interest as the system composed of the two planets. This}
 resonance locked system interacts 
with the inner disk through torques proportional to
$M_J^2$, at Jupiter's inner Lindblad resonances (ILR), 
whereas it interacts with the outer disk through torques proportional
to $M_S^2$ at Saturn's outer Lindblad resonances (OLR). 
Now Saturn's ILR fall in Jupiter's gap and Jupiter's
OLR fall in Saturn's gap so their effect is weakened compared to the situation
where Jupiter is alone. As $M_J^2/M_S^2\sim 10$,
the torque imbalance does not favor an inwards migration as strongly as in a one planet case,
and can even lead to a positive differential
Lindblad torque on the two planet system.
{Actually one can estimate what the maximum mass ratio of the outer
planet to the inner one should be
to get a migration reversal, if one neglects the Inner Lindblad torque  on the outer planet and the
Outer Lindblad torque on the inner planet.
One can show, with conservative assumptions, that the reversal occurs if $\mu_S/\mu_J < q_c$, with
$q_c\sim 0.6$ (Masset \& Snellgrove 2000).
}

\section{Discussion}

We have performed a series of ``restart runs'' in order
to check for a variety of behaviors. These runs are restarted once
the
planets are locked in resonance, and only one parameter is changed at
one time. These runs have enabled us to:\\
(i) Confirm that the differential Lindblad torque on the system is
positive, with a restart run in which the aspect ratio is changed (see Masset \& Snellgrove
2000 for further details).\\
(ii) Check that the accretion onto Jupiter does not change the
outwards migration rate, except in a maximally accreting case.\\
(iii) Check that the situation presented here is a steady state
one. Indeed, the common gap turns out to be 'permeable' to the flow of
material from the outer disk to the inner one, since in all our runs
the mass flow through the gap adjusts to $3\pi\nu\Sigma_0+2\pi
r_s\dot r_s\Sigma_0$,
where $\Sigma_0$ is the disk unperturbed surface density and $r_s$ is
the radius of the separatrix between the last circulating streamline
in the outer disk and Saturn's coorbital horseshoe region.

Since the situation presented here is a steady state one, the outwards
migration can be sustained as long as the planets are trapped into
resonance. This is not necessarily the case: it will be 
presented elsewhere in greater
detail how Saturn can be moved apart from the $2:3$ resonance with
Jupiter and be sent further out in the disk on a short timescale (a
few tens of orbits). On the other hand, it should be noted that if 
the planets are still locked in resonance when the gas disappears, the
eccentricities will not be confined anymore to small values by the
coorbital material, and the system is likely to become unstable. This
could result in the ejection of one planet, while the other one would
remain on an eccentric orbit.

\end{document}